\documentclass[preprint]{sig-alternate}

\usepackage{amsfonts}
\usepackage{url}
\usepackage{fancyvrb}
\usepackage{booktabs}
\usepackage{multirow}
\usepackage{tabularx}
\usepackage[normalem]{ulem}
\usepackage{tikz}
\usetikzlibrary{arrows}
\usetikzlibrary{calc}
\usetikzlibrary{fit}
\usetikzlibrary{patterns}
\usetikzlibrary{shadows}
\usetikzlibrary{shapes.symbols}
\usetikzlibrary{shapes.geometric}
\usetikzlibrary{shapes.multipart}
\usetikzlibrary{shapes.misc}
\usetikzlibrary{shapes.arrows}
\usetikzlibrary{decorations.pathmorphing}
\pgfdeclarelayer{foreground}
\pgfdeclarelayer{background}
\pgfdeclarelayer{middle}
\pgfsetlayers{background,middle,main,foreground} 

\usepackage[utf8]{inputenc}

\usepackage{color,pdfcolmk}

\newcommand{\F}{Figure}
\newcommand{\T}{Table}
\renewcommand{\S}{Section}

\newcommand{\etal}{\textit{et al.}}

\newcommand{\hdbg}{\textsc{HyperDbg}}

\newfont{\emailfnt}{phvr8t at 11.5pt}
\renewcommand{\email}[1]{{{\emailfnt{#1}}}}       

\title{Dynamic and Transparent Analysis of Commodity Production Systems}
\author{
  \begin{tabular}{cccc}
    \hspace{0.75em} Aristide Fattori$^\dag$ &
    Roberto Paleari$^\dag$ &
    Lorenzo Martignoni$^\ddag$ &
    Mattia Monga$^\dag$ \\
    \multicolumn{2}{c}{\email{\{aristide,roberto\}@security.dico.unimi.it}} & 
    \email{lorenzo.martignoni@uniud.it} &
    \email{mattia.monga@unimi.it} \\
  \end{tabular} \\
  \begin{tabular}{cc}
    & \\
    \affaddr{Dip. di Informatica e Comunicazione$^\dag$} & 
    \affaddr{Dip. di Fisica$^\ddag$} \\
    \affaddr{Universit\`a degli Studi di Milano} & 
    \affaddr{Universit\`a degli Studi di Udine}\\
    \affaddr{I-20135 Milan, Italy} & 
    \affaddr{I-33100 Udine, Italy}
  \end{tabular}
  \\
}

\begin{document}
\toappear{To appear in the $25^{th}$ IEEE/ACM International Conference
  on Automated Software Engineering, Antwerp, Belgium, 20-24 September
  2010} \clubpenalty=10000 \widowpenalty=10000

\maketitle{}

\begin{abstract}
  We propose a framework that provides a programming interface to perform complex
dynamic system-level analyses of deployed production systems. By leveraging
hardware support for virtualization available nowadays on all commodity
machines, our framework is completely transparent to the system under analysis
and it guarantees isolation of the analysis tools running on its top.  Thus,
the internals of the kernel of the running system needs not to be modified and
the whole platform runs unaware of the framework. Moreover, errors in the
analysis tools do not affect the running system and the framework.  This is
accomplished by installing a minimalistic virtual machine monitor and migrating
the system, as it runs, into a virtual machine. In order to demonstrate the
potentials of our framework we developed an interactive kernel debugger,
nicknamed \hdbg. \hdbg{} can be used to debug any critical kernel component,
and even to single step the execution of exception and interrupt handlers.


\end{abstract}

\category{D.2.5}{Software Engineering}{Testing and
  Debugging}[Debugging aids, Monitors, Tracing]
\category{D.4.9}{Operating Systems}{Systems Programs and Utilities}

\terms{Verification}

\keywords{hardware virtualization, debugging, system analysis}

\section{Introduction}
\label{sec:intro}

Operating systems are peculiar and very complex pieces of software whose
internals are critically vital for a system: a failure, or a bottleneck, in any
of their parts can lead to catastrophic consequences. Therefore, special care
is needed to develop, analyze, test, and profile them. To simplify their task,
developers and analysts rely on a large variety of tools and analysis
techniques. Some of them are specific for studying static properties of the
operating system, while others are more specific for studying dynamic
properties. In particular, the latter class of tools and techniques is nowadays
very popular among kernel developers and analysts because it allows them to
collect the information very quickly, while hiding many of the intricacies of
the kernel, and can even be used on running production systems.

Existing approaches for dynamic analysis of operating systems (\textit{e.g.,} debugging,
profiling, and tracing) can be roughly classified in two groups: kernel-based
and VMM-based. The approach taken by the first group is to include some
component into the kernel in order to intercept all the events of interest
(\textit{e.g.,} the creation of a new process, the execution of a system call, and the
execution of a kernel function) and to execute a specific action when such
events
occur~\cite{cantrill04:dtrace,ltt,windbg, softice,tamches01:kerninst}. This
solution requires the installation of specific hooks in the kernel to monitor
run-time events and it might be very difficult to apply to operating systems
that do not natively offer facilities for dynamic analysis, especially when the
source code is not available. The approach taken by the second group is to run
the kernel and user-space applications in a virtual machine and to intercept,
and respond to, the events of interest from the virtual machine monitor
(VMM)~\cite{garfinkel03:vmi}. Although this approach guarantees transparency
and has a loose dependency on the operating system internals, it cannot be used
in all the settings, since it implies that the system must be run as a guest of
a virtual machine and production systems not running in virtual machines cannot
be analyzed. Moreover, VMM-based solutions typically virtualize hardware
devices, to allow multiple guests to share the same physical peripherals. This
makes software virtualization approaches unsuitable to assist the analysis of
components that need to interact directly with the underlying hardware.

In this paper we propose a framework that brings together the advantages of
both approaches: it can be used on commodity production systems (\textit{i.e.,}
off-the-shelf products, whose source code or debugging symbols are not
necessarily available), since it does
not require to instrument the system under test, and it is able to inspect
systems running on real hardware, since it does not require an emulation
container. Similarly to existing frameworks, the analyses that can be built on
top of our framework include profiling and tracing of the kernel and user-space
applications, interactive debugging, or even extension of system
features. However, differently from existing frameworks, ours is \emph{fully
  dynamic}, \emph{transparent}, \emph{loosely dependent on the operating
  system}, and \emph{fault-tolerant} with respect to possible defects in the
analysis code.
First, our framework does not require recompilation or rebooting of the target
system. Thus, it can be used to analyze any running production system,
including commodity operating systems lacking native support for
instrumentation and systems not running in virtual machines. Second, the
framework is not invasive, since analyses can be performed on a virtually
unmodified system: as explained in the paper, only a minimal driver needs to be
installed and no parts of the kernel are patched in any way. Moreover, since
the framework itself is not accessible from the target system, its code cannot
be detected by malicious code or unwittingly influence buggy operating system
components. Thus, the infrastructure can be applied to any operating system, as
the majority of the facilities it supports are completely OS-independent, and
the only OS-dependent functionalities are just provided to ease the development
of analysis tools. Finally, the framework is fault-tolerant, as it guarantees
that a defect in an analysis tool built on top of it do not damage the framework itself
nor the analyzed system.


Our framework leverages hardware extensions for virtualization available on
commodity x86 CPUs~\cite{svm,neiger06}. Hardware-support for virtualization
allows the development of virtual machine monitors that are very efficient,
completely transparent, and non invasive to the systems running in the virtual
machine. To overcome the major limitation of traditional VMM-based approaches
(\textit{i.e.,} the impossibility to analyze productions systems not running in a
virtual machine), our framework exploits a feature of the hardware that allows
to install a virtual machine monitor and to \emph{migrate a running system into
  a virtual machine}. When the analysis is completed, the original mode of
operation of the system can be restored. Practically speaking, our framework is
a minimalistic virtual machine monitor acting as a broker between the analyzed
system and the analysis tool. The framework abstracts low-level events
occurring in the analyzed system into high-level events and guarantees
fault-tolerance by relying on the hardware to run the analysis tool in a
isolated execution environment.



To demonstrate the potentials of our framework we have developed an interactive
kernel debugger, nicknamed \hdbg, constructed entirely using the programming
interface exposed by our infrastructure. \hdbg{} adds live and interactive
debugging support to Microsoft Windows XP, so far only possible using very
invasive tools, like Syser~\cite{syser}, or traditional VMM-based
debuggers. \hdbg{} can be used to debug any component of the Windows kernel,
including interrupt/exception handlers, device drivers, and even supports
single instruction stepping. Being completely separated from the debuggee,
\hdbg{} is transparent to the analyzed system and can be even used to analyze
protected and malicious code.

In summary, the paper makes the following contributions.
\begin{enumerate}
\item We propose a framework to perform complex dynamic system-level analyses
  of commodity production systems. Compared to existing frameworks, the one we
  propose guarantees transparency, efficiency, and does not require the target
  system to be already installed on a virtual machine.
\item We implemented our framework in an experimental prototype for Microsoft
  Windows XP.
\item We describe the design and the implementation of \hdbg, a kernel-level
  interactive debugger built on top our framework. 
\end{enumerate}

Both the analysis framework and \hdbg{} are available at
\url{http://security.dico.unimi.it/hyperdbg/} and is released under the terms
and conditions of the GPL (v3.0) license.


\section{Related Work}
\label{sec:related}

The framework proposed in this paper shares many similarities with frameworks
and techniques extensively explored in the past. However, by exploiting recent
facilities available of modern Intel x86 CPUs, our framework is able to combine
and to offer simultaneously the main benefits introduced by previous research
work.

\paragraph{Dynamic Kernel Instrumentation}
DTrace is a facility included into the Solaris kernel that allows the dynamic
instrumentation of production systems~\cite{cantrill04:dtrace}. The key points
of DTrace are efficiency and flexibility. First, the instrumentation framework
itself introduces no overhead. Second, the framework provides tens of thousands
of instrumentation points, and the actions to be taken can be expressed in
terms of a high-level control language, that also includes a number of
mechanisms to guarantee run-time safety.
Similarly, KernInst is a dynamic instrumentation framework for commodity
kernels~\cite{tamches01:kerninst}. KernInst has been developed mainly to gather
information about the performances of a running kernel, but it has also been
employed for run-time kernel optimization. Differently from DTrace, KernInst
does not provide any mechanism for run-time safety of the instrumentation
routines. Unfortunately, the aforementioned approaches are not transparent, as
they require direct modifications of the operating system kernel,
achieved by loading a kernel-mode module. Moreover,
none of them is OS-independent, and they and cannot be applied to closed-source
operating systems. Our framework does not suffer these limitations since it
can instrument the kernel without modifying it and does not rely on any
facility offered by the kernel. 

\paragraph{Kernel-level Debugging}

Several efforts have been made to develop efficient and reliable kernel-level
debuggers. Indeed, these applications are essential for many activities, such
as the development of device drivers. One of the first and most widely used
kernel-level debuggers that targeted the Microsoft Windows operating system was
SoftICE~\cite{softice}, but today the project has been discontinued. However,
both commercial~\cite{syser} and open-source~\cite{rr0d} alternatives to
SoftICE appeared. Modern versions of Windows already include a kernel debugging
subsystem~\cite{windbg}. Unfortunately, to exploit the full capabilities of
Microsoft's debugging infrastructure, the host being debugged must be
physically linked (\textit{e.g.,} by means of a serial cable) with another machine. All
these approaches share a common factor: to debug kernel-level code, they
leverage another kernel-level module. Obviously, that is like a dog chasing its
tail. The framework proposed in this paper does not require any kernel support
nor to modify the kernel to add the missing support at run-time.

\paragraph{Frameworks Based on Virtual Machines}

Instead of relying on a kernel-level module to monitor other kernel code, an
alternative approach consists of running the target code inside a virtual
machine, and to perform the required analyses from the
outside~\cite{garfinkel03:vmi}. In~\cite{king05:osdebug,xu07:retrace,dunlap02:revirt}
the authors propose virtual machines with execution replaying capabilities: a
user can move forward and backwards through the execution history of the whole
system, both for debugging and for understanding how a hacker intrusion took
place. Finally, in~\cite{chow08:aftersight} Chow \etal{} propose Aftersight, a
system that decouples execution recording from execution trace analysis, thus
reducing the overhead suffered by the system where the guest operating system
is run. Nowadays, Aftersight is part of the VMware platform, and other
mainstream commercial products provide similar capabilities. The framework
proposed in this paper can provide these functionalities even on systems not
running in any virtual machine. 

\paragraph{Aspect-oriented Programming}

Aspect-oriented programming is a paradigm that promises to increase modularity
by encapsulating cross-cutting concerns into separated code units, called
``aspects'', whose ``advice'' code is woven into the system automatically, by
specifying the properties of the join-points. AspectC is an aspect-oriented
framework that is used to customize (at compile-time) operating system kernels
\cite{coady01:kernel, mahrenholz02:pure, mahrenholz02:instrument}. More dynamic
approaches have been proposed: for example TOSKANA provides \emph{before},
\emph{after} and \emph{around} advices for in-kernel functions and supports the
implementation of aspects themselves as dynamically exchangeable kernel
modules~\cite{engel06:toskana}. The framework proposed in this paper allows to
achieve the same goal while being transparent and fault-tolerant.


\section{Overview of the Framework}
\label{sec:overview}

\begin{figure}[tbp]
  \centering
  \begin{tikzpicture}
  \tikzstyle{block} = [draw, rounded corners, fill=white, drop shadow, font=\scriptsize, 
                       inner xsep=4pt, minimum height=0.7cm];
  \tikzstyle{kernelline} = [dashed, font=\scriptsize\it];
  \tikzstyle{vmxline}    = [thick,  font=\scriptsize\it];
  \tikzstyle{arrow}      = [->, >=stealth, rounded corners, font=\scriptsize\it];
  \tikzstyle{bigarrow}   = [>=stealth, ->, line width=5pt, color=gray!50, rounded corners, font=\it];

  \begin{scope}[xshift=-4.6cm]
    \node[block, minimum width=3.4cm] at (0,0) (os) {Operating system kernel};

    \draw[kernelline] ($(os.north east) + (0,0.4)$) -- ($(os.north west) + (0,0.4)$);
    \node[kernelline, anchor=east] at ($(os.north east) + (-0.20,0.54)$) {User mode};
    \node[kernelline, anchor=east] at ($(os.north east) + (-0.20,0.26)$) {Kernel mode};

    \node[above of=os, node distance=1.6cm] (pivot) {};
    \node[block, minimum width=1.5cm, right of=pivot, node distance=1cm] (p1) {
      \begin{tabular}{c}
        User\\process
      \end{tabular}
    };
    \node[block, minimum width=1.5cm, left of=p1, anchor=east, node distance=1cm] (p2) {
      \begin{tabular}{c}
        User\\process
      \end{tabular}
    };

    \node[fit=(p1) (p2) (os)] (fit1) {};
  \end{scope}

  \begin{scope}
    \node[block, minimum width=3.4cm] at (0,0) (os) {Operating system kernel};

    \draw[kernelline] ($(os.north east) + (0,0.4)$) -- ($(os.north west) + (0,0.4)$);
    \node[kernelline, anchor=east] at ($(os.north east) + (-0.20,0.54)$) {User mode};
    \node[kernelline, anchor=east] at ($(os.north east) + (-0.20,0.26)$) {Kernel mode};

    \node[above of=os, node distance=1.6cm] (pivot) {};
    \node[block, minimum width=1.5cm, right of=pivot, node distance=1cm] (p1) {
      \begin{tabular}{c}
        User\\process
      \end{tabular}
    };
    \node[block, minimum width=1.5cm, left of=p1, anchor=east, node distance=1cm] (p2) {
      \begin{tabular}{c}
        User\\process
      \end{tabular}
    };

    \draw[vmxline] ($(os.south east) + (0,-0.4)$) -- ($(os.south west) + (0,-0.4)$);
    \node[vmxline, anchor=east] at ($(os.south east) + (0,-0.27)$) {Non-root mode};
    \node[vmxline, anchor=east] at ($(os.south east) + (0,-0.57)$) {Root mode};

    \node[block, below of=os, node distance=3.1cm, minimum width=2.4cm] (core) {Framework};
    \node[block, minimum width=1.2cm, below of=p1, node distance=3.15cm, inner sep=0] (a1) {
      \begin{tabular}{c}
        Analysis\\tool
      \end{tabular}
    };


    \begin{pgfonlayer}{foreground}
      \draw[arrow,-] (p2.south) ++(-0.2cm, 0cm) -- ++(0cm, -0.85cm);
      \draw[arrow,-,dashed,gray] (p2.south) ++(-0.2cm, -0.85cm) -- ++(0cm, -0.7cm);
      \draw[arrow] (p2.south) ++(-0.2cm, -1.55cm) -- node[rotate=90, above=-0.1cm, pos=0.6] {Exit} ++(0cm, -2.4cm);
      \draw[arrow,-] ($(os.south) + (-1.2cm, 0cm)$) -- ++(0cm, -1.0cm) -- ++(0.22cm, 0cm);

      \draw[arrow,<-] (p2.south) ++(+0.2cm, 0cm) -- ++(0cm, -0.85cm);
      \draw[arrow,-,dashed,gray] (p2.south) ++(+0.2cm, -0.85cm) -- ++(0cm, -0.7cm);
      \draw[arrow,-] (p2.south) ++(+0.2cm, -1.55cm) -- node[rotate=90, above=-0.1cm, pos=0.6] {Inspect} ++(0cm, -2.4cm);
      \draw[arrow,<-] ($(os.south) + (-0.365cm, 0cm)$) -- ++(0cm, -1.0cm) -- ++(-0.22cm, 0cm);

      \draw[arrow,<->] (a1.south) ++(0.5cm,0cm) |- (core.east);
    \end{pgfonlayer}
  \end{scope}

  \draw[bigarrow]    (fit1.south) ++(-0.4cm, -0.1cm) -- ++(0cm,  -2.0cm) -- ++(3.0cm, 0cm) 
                     node[black,below,pos=0.47] {Install};

  \draw[bigarrow,<-] (fit1.south) ++(0.4cm, -0.1cm) -- ++(0cm, -1.0cm) -- ++(2.2cm, 0cm)
                     node[black,below,pos=0.33] {Remove};
\end{tikzpicture}

  \caption{Overview of the framework}
  \label{fig:arch}
\end{figure}
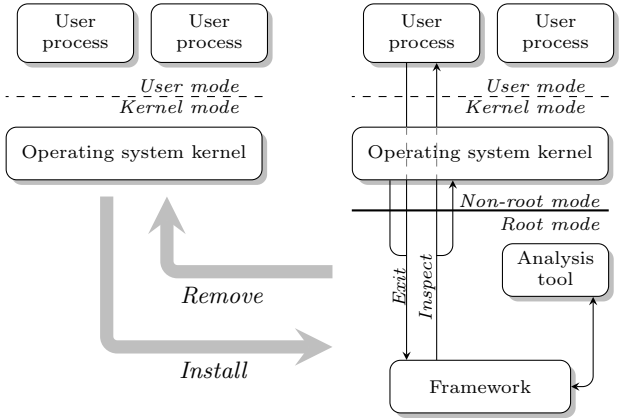

\F~\ref{fig:arch} depicts the architecture of our framework, 
the installation and removal processes, and the migration of the operating
system and its applications into a virtual machine. Our framework consists of a
virtual machine monitor (VMM for short) that provides a programming interface
for the development of system-level analysis tools. As in traditional VMM-based
analysis approaches, the analysis tool is run within the VMM and thus
completely transparent to guests of the virtual machine. However, compared to
traditional VMM-based ones, ours does not require the system to be already
running inside any virtual machine. To achieve this goal, our framework
leverages hardware extensions for virtualization available on all modern x86
CPUs~\cite{neiger06,svm} (which are unused in the majority of the
deployments). In short, these extensions augment the instruction set
architecture with two new modes of operation: \emph{VMX root mode} and
\emph{VMX non-root mode}\footnote{VMX (non-) root mode is the terminology used
  by Intel; AMD adopts a different terminology.}. These new modes of operation
separate logically the virtual machine monitor from a guest without having to
modify the latter. More precisely, we exploit a particular feature of these
extensions that allows for \emph{late launching of VMX modes}. Late launching
of VMX modes permits to install a virtual machine monitor even if the system
has already been bootstrapped. In other words, late launching allows to migrate
(temporarily) a running operating system in a virtual machine, and to analyze
and control the execution of the system from the monitor. Through the rest of
the paper, we use the term ``guest'' to refer to the system under analysis that
has been migrated into a virtual machine.

Practically speaking, the running operating system is not migrated anywhere and
not touched at all.
Rather, by launching VMX modes, the execution environment is extended with the
two aforementioned operating modes; the running operating system is then
associated with non-root mode, while the VMM is associated with root mode.
Thus, in all respects, the operating system and its applications become a guest
of our special virtual machine.  Following the same principle, the VMM can be
unloaded, and the original mode of execution of the operating system restored,
by simply disabling VMX modes. After the launch of the VMX modes, the execution
of the guest can continue exactly as before, even in terms of interactions with
the underlying hardware devices. However, during its execution, the guest might
be interrupted by an \emph{exit} to root mode. Like hardware exceptions, exits are
events that block the execution of the guest, switch from non-root mode to root
mode, and transfer the control to the VMM. Differently from exceptions, the set
of events triggering exits to root mode can be configured dynamically by the
VMM. A routine of the VMM handles the exit and eventually \emph{enters}
non-root mode to resume the execution of the guest. Being executed at the
highest privilege level, the routine handling the exit has complete read/write
control of the state of the guest system (of both memory and CPU registers).


The framework itself does not perform any analysis. It is only responsible for
handling a small set of exits to control all accesses to the memory management
unit of the CPU, to prevent the guest from accessing the physical memory
locations holding the code and the data of the framework. On the other hand,
the framework provides a flexible API to develop tools to perform sophisticated
analyses of both kernel and user code running in the guest. Using the
functionalities provided through the API, the tool can request the framework to
monitor certain events that might occur during the execution of the guest; when
such events occur, it can inspect, and even manipulate, the state of the
guest. The events that can be monitored include, but are not limited to, system
call invocations, function calls, context switches and I/O
operations. Practically speaking, events are monitored through exits to root
mode. Thus, a request of the analysis tool to monitor a certain high-level
event (\textit{e.g.,} the execution of a system call) is translated by the API of the
framework into a sequence of low-level operations that guarantee that all the
occurrences of such event in the guest trigger an exit to root mode. Similarly,
the framework translates the exit into a higher-level event and notifies the
occurrence of the event to the analysis tool. Once notified, the tool can
recover information about the event (\textit{e.g.,} arguments and return value of a
system call), using the inspection functionalities offered by the API.

An important requirement for the analysis of production systems is that
analysis tools must not interfere with the correct execution of the guest. This
is particularly important for faults and deadlocks that might occur in the
analysis tool. The approach we adopt is to run the tool in a
less privileged execution environment, isolated from the analyzed system and
from the framework. The tool can interact with the guest only through the API
exposed by the framework. This approach guarantees the framework the ability to
intercept any fault occurring in the tool, to mediate all accesses to the
analyzed system (and to prevent write accesses), and to terminate the tool in
case of deadlocks or other anomalous situations.



\section{Design and implementation}
\label{sec:details}


\begin{figure}[tbp]
  \centering
  \begin{tikzpicture}
  \tikzstyle{block} = [draw, rounded corners, fill=white, drop shadow, font=\scriptsize, 
                       inner xsep=4pt, minimum height=0.7cm];
  \tikzstyle{module} = [fill=gray!20, rounded corners, draw, inner ysep=13, inner xsep=8];
  \tikzstyle{kernelline} = [dashed, font=\scriptsize\it];
  \tikzstyle{vmxline}    = [thick,  font=\scriptsize\it];
  \tikzstyle{hwline}    = [very thick,  font=\scriptsize\it];
  \tikzstyle{arrow}      = [->, >=stealth, rounded corners, font=\scriptsize\it];
  \tikzstyle{device}     = [block, text width=1.1cm, text centered];

  \node[block] at (2.0,-2.6) (eg) {Event gate};
  \node[block, right of=eg, node distance=1.9cm] (tg) {Trap gate};

  \node[block, right of=tg, node distance=3.40cm] (ag) {API};
  \node[block, above of=tg, node distance=2.75cm] (at) {Analysis tool};

  \draw[arrow, <-] (eg) -- node[above,sloped,pos=0.6] {1. Exit} ++(0cm,4.1cm);
  \draw[arrow] ($(eg.north) + (0.5cm,0cm)$) |- node[above,rotate=90,pos=0.45] {2. Notification} (at);
  \draw[arrow] (at) -- node[right=-0.2cm,pos=0.45] {\begin{tabular}{l}3. API call\\6. Exception\end{tabular}} (tg);
  \draw[arrow] (tg) -- node[above, sloped] {4. API request} (ag);
  \draw[arrow] (ag) -- node[above=0.1cm, sloped, pos=0.58, fill=white, inner sep=1pt] {4a. Inspect/manipulate} ++(0cm,4.1cm);

  \draw[arrow] 
  ($(ag.south) + (-0.25cm,0cm)$) -- ++(0cm,-0.8cm) -| 
  node[above,pos=0.25, fill=white, inner sep=0pt] {4b. Request event notification}
  ($(eg.south) + (0.25cm,0cm)$);

  \draw[arrow,<-]
  ($(ag.south) + (0.25cm,0cm)$) -- ++(0cm,-1.3cm) -| 
  node[above,pos=0.25, fill=white, inner sep=0pt] {5. Recover information about events} 
  ($(eg.south) + (-0.25cm,0cm)$);

  \begin{scope}[xshift=0.4cm]
    \node[vmxline, anchor=west, fill=white, inner sep=1pt] at (-0.10, 1.1) {Non-root mode};
    \node[vmxline, anchor=west, fill=white, inner sep=1pt] at (-0.10, 0.7) {Root mode};

    \node[kernelline, anchor=west] at (-0.15,-0.90) {User mode};
    \node[kernelline, anchor=west] at (-0.15,-1.30) {Kernel mode};

    \node[hwline, anchor=west] at (-0.15,-4.7) {Hardware};

    \begin{pgfonlayer}{background}
      \draw[vmxline] (-0.1,0.9) -- ++(8.2,0);
      \draw[kernelline] (-0.1,-1.1) -- ++(8.2,0);
      \draw[hwline] (-0.1,-4.5) -- ++(8.2,0);
    \end{pgfonlayer}
  \end{scope}

  \begin{pgfonlayer}{background}
    \node[module, fit=(eg) (tg) (ag)] (core) {};
  \end{pgfonlayer}

  \node[font=\scriptsize,inner sep=1pt,fill=gray!20,anchor=north east] at ($(core.north east)+(-0.05cm,-0.05cm)$) {Framework};

  \node[device] at ($(tg.south) + (0cm,-2.3cm)$) (pit) {Timer};
  \node[device, right of=pit,  node distance=1.6cm] (dev1) {\color{gray} Disk};
  \node[device, right of=dev1, node distance=1.6cm] (dev2) {\color{gray} Network};
  \node[device, left  of=pit,  node distance=1.6cm] (dev3) {\color{gray} Video};

  \begin{pgfonlayer}{background}
    \draw[arrow] (pit) -- node[right, pos=0.10] {7. Interrupt} (tg);
  \end{pgfonlayer}
\end{tikzpicture}

  \caption{A close-up of the framework}
  \label{fig:framework}
\end{figure}
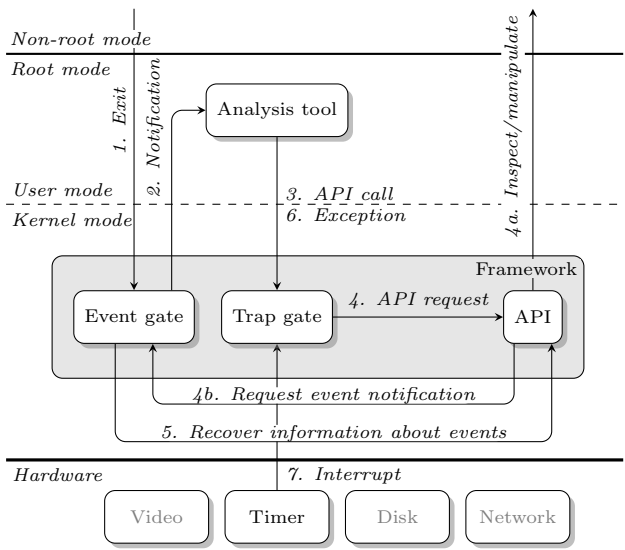

\F~\ref{fig:framework} shows a more detailed view of the architecture of our
framework. Intuitively, this architecture is very similar to that of
traditional operating systems: the framework plays the role of the kernel and
the analysis tool plays the role of a user-space application. As will become
clear later, this architecture prevents buggy analysis tools from compromising
the guest system and the framework. The separation between these two parts
is made possible by the fact that, when VMX is enabled, root and
non-root modes offer two fully-featured execution environments. Thus, like the
guest running in non-root mode, the framework running in root mode can rely on
privilege separation to isolate the analysis tool and can handle independently
interrupts and exceptions that might occur while executing in root mode.

When an exit to root mode interrupts the execution of the guest, the event is
delivered to the \emph{event gate} (step 1 in \F~\ref{fig:framework}). The
event gate is responsible for abstracting low-level events into higher-level
ones, and to notify the analysis tool if the latter has requested to do so
(step 2). On startup the analysis tool requests the framework to be notified of
certain events (not shown in the figure). The tool can use the API provided by
framework to query extra information about the event (\textit{e.g.,} the content of the
stack location storing one of the arguments of a function). Since the tool is
isolated from the framework, API functions are invoked through software
interrupts. Thus, requests coming from the analysis tool are received by the
\emph{trap gate} (step 3), then forwarded to the component implementing the API
(step 4). The tool can perform two types of API calls: (step 4a) to inspect or
manipulate the state of the guest, and (step 4b) to control event notifications
(\textit{e.g.,} enable or disable the notification of certain events). Note that the
component implementing the API is also used by the framework itself (step 5) to
recover extra information about events (\textit{e.g.,} the return address of a function
stored in the stack). The trap gate also serves the purpose of detecting
exceptions (\textit{e.g.,} page faults) that might occur during the execution of the
analysis tool. If the trap gate intercepts an exception (step 6), it terminates
the faulty tool and unloads the framework, to resume the normal operation mode
of the system. Finally, the trap gate is also used to handle timer interrupts
(step 7), that, as will be discussed in \S~\ref{sec:isolation}, are employed to
enforce a time-bound on the execution of the tool.

\begin{table*}[htbp]
  \centering
  {
    \scriptsize
    \begin{tabular}{lllc}
      \toprule
      \textbf{Event} & \textbf{Description} & \textbf{Arguments}\\
      \midrule
      \texttt{ProcessSwitch}    & Context (process) switch & --- \\
      \texttt{Exception} & Execution & Exception vector, faulty instruction,
      error code \\
      \texttt{Interrupt} & Hardware or software interrupt & Interrupt vector,
      requesting instruction \\
      \texttt{BreakpointHit}    & Execution breakpoint & Breakpoint address \\
      \texttt{WatchpointHit}    & Watchpoint on data read/write & Watchpoint
      address, access type, hitting instruction \\

      \texttt{FunctionEntry}    & Function call & Function name/address, caller/return address \\
      \texttt{FunctionExit}      & Return from function & Function
      name/address, return address \\
      \texttt{SyscallEntry}  & System call invocation & System call number,
      caller/return address \\
      \texttt{SyscallExit}   & Return from system call & System call number,
      return address\\
      \texttt{IOOperationPort}  & I/O operation throught hardware port & Port number, access type \\
      \texttt{IOOperationMmap}  & Memory-mapped I/O operation & Memory address, access type \\
      \bottomrule
    \end{tabular}
  }
  \caption{Events traceable using our framework and corresponding arguments
    (the argument that represents the current process is omitted, as it is common to all the events)}
  \label{tab:events}
\end{table*}

The functionalities provided by the API of the framework can be classified into
two classes: \emph{execution and I/O tracing} and \emph{state inspection and
  manipulation}. The following paragraphs describe briefly the API. More
details are given in Sections \ref{sec:exec} and \ref{sec:mmu}.

Execution and I/O tracing facilities allow a tool to intercept the occurrence
in the analyzed system of certain events and certain I/O operations
respectively. \T~\ref{tab:events} reports the main types of events that can be traced. For each event, the table also reports the arguments
associated to the event; arguments are information about the events most
commonly used in tools. For example, the events \texttt{FunctionEntry} and
\texttt{SyscallEntry} are used to trace functions and system calls
respectively. The arguments associated to the \texttt{FunctionEntry} event are
the address (or the name) of the function called, the caller and the return
address. Another example is the \texttt{ProcessSwitch} event that can be used
to trace context switches between processes (not threads). From the point of
view of the analysis tool all the events are handled in the same way: the tool
can subscribe to any event and, when the event occurs, can inspect its
arguments and take the proper actions. However, at the framework-level, certain
events are different from other ones. Indeed, some of them (\textit{e.g.,} context
switches between processes) can be traced directly by the hardware. That is,
the event triggering the exit corresponds exactly to the event being
traced. Other events instead (\textit{e.g.,} function calls and returns) cannot be
traced directly by the hardware. In all these cases the framework relies on
other low-level events to trace the execution and then abstract exiting
low-level events into higher-level ones, meaningful for the analysis tool.

Arguments can optionally be used as conditions, to limit the tracing to a
subset of all the events. Conditions on events serve two purposes. First,
conditions allow to simplify the analysis tools, since events that do not match
the requested conditions are discarded by the framework and thus do not need to
be handled by the tool. Second, some conditions allow preemptive filtering of
the events. In other words, the framework configures \emph{a priori} which
events trigger an exit, instead of filtering out exits caused by uninteresting
events. For example, in the case of the \texttt{IOOperationPort} event,
preemptive filtering means to configure the CPU such that only I/O operations
involving a specific I/O port trigger an exit. This feature is very important
to minimize the number of exits and thus the overall overhead.

State inspection and manipulation primitives can be used by the tool to access
the state of the guest, in order to extract more detailed information about
events or other data useful for the analysis. For example, these primitives
allow to extract the arguments of an invoked function, or to inspect the
internal structures of the guest operating system. Note that, by default, write
access to guest state is not granted to a tool.  If necessary, such permission
can be enabled at compile-time. Obviously, in this case the framework cannot
protect the state of the guest from dangerous modifications.

\subsection{Framework and Analysis Tool Loading}
\label{sec:late}

The framework and the analysis tool are loaded by a minimal kernel driver. This
is unavoidable since the operations we need to perform to load the framework
require maximum privileges and can be performed only by the kernel of the
operating system. The driver, however, is indeed very simple and we put extreme
care in avoiding any interference with the kernel. Moreover, since once loaded
the framework is completely invisible to the system, we unload the driver immediately
as soon as the framework has been installed.

When VMX modes are enabled, a special VMX data structure (VMCS in Intel
terminology) is made accessible initially to the loader, and subsequently, when
the loading is completed, only to the framework. This data structure stores the
\emph{host state}, \emph{guest state}, and the \emph{execution control fields}.
The host state stores the state of the processor that is loaded on exits to
root mode, and consists of the state of all the registers of the CPU (except
for general purpose registers). Similarly, the guest state stores the state of
the processor that is loaded on entries to non-root mode. The guest state is
updated automatically at every exit, such that the subsequent entry to non-root
mode will resume the execution from the same point. The execution control
fields allow a fine-grained specification of which events should trigger an
exit to root mode.

The task of the loader is to enable VMX modes and to configure the VMX data
structure such that the execution of the operating system and user-space
applications continue to run in non-root mode, while the framework and the
analysis tool are executed in root mode. Moreover, the loader has to configure
the CPU such that all the events necessary for the tool to trace the execution
of the system trigger exits to root mode. When the initialization is completed,
the driver unloads itself and resumes the execution of the system.

\paragraph{Guest State Configuration}

The \emph{guest state} is initialized to the current state of the system. In
this way, when the virtual machine is launched and execution enters non-root
mode, the guest operating system will resume its execution as if nothing
happened. A tricky problem when initializing non-root mode concerns the
management of the memory. More precisely, we must prevent the newly created
guest to use and access the physical memory frames allocated to the framework
and to the tool. Otherwise, the guest could detect and even corrupt the
framework. Most recent CPUs provide hardware facilities for memory
virtualization (\textit{e.g.,} Intel Extended Page Table extension). If these facilities
are not available, memory virtualization must be implemented entirely via
software. Briefly, software memory virtualization consists of intercepting all
guest operations to manipulate the page table (the data structure the CPU uses
for virtual-to-physical address translation) and in ensuring that none of the
physical frames allocated to the framework and to the analysis tool are mapped
into the guest. In case the guest tries to map a reserved physical frame, the
framework assigns the guest a different one and masquerades the difference.

\paragraph{Host State Configuration}

The host state is initialized as follows. The CPU is configured to use, when in
root mode, a dedicated address space and a dedicated interrupt descriptor table
(IDT). This configuration simplifies the separation of the analyzed system from
the framework and allows to detect and handle interrupts and exceptions that
occur in root mode. Differently from the address of the entry point of non-root
mode, which is updated at every exit to allow to resume execution of the guest
from where it was interrupted, the address of the entry point of root mode is
fixed. The entry point is set to the address of the routine that takes care of
dispatching an exit event to the appropriate handler and that in turn might
notify the analysis tool (\textit{i.e.,} the entry point of the event gate).

\paragraph{Execution Control Fields Configuration}

To reduce the run-time overhead suffered by the guest system, the execution
control fields are configured to minimize the number of events that trigger an
exit to root mode. When the tool is initialized, it specifies which
events must be intercepted. Subsequently, in response to the invocation of API
functions, the configuration of the execution control fields can be altered to
intercept additional events or to ignore other ones.

\subsection{Execution Tracing}
\label{sec:exec}

\begin{table}[tbp]
  \centering
  {
    \scriptsize
    \begin{tabular}{llc}
      \toprule
      \multirow{2}{*}{\textbf{Event}} & \multirow{2}{*}{\textbf{Exit cause}} & \textbf{Native} \\
      & & \textbf{exit} \\
      \midrule
      \texttt{ProcessSwitch}    & Change of page table address  & $\surd$ \\
      \texttt{Exception} & Exception & $\surd$ \\
      \texttt{Interrupt} & Interrupt & $\surd$ \\
      \texttt{BreakpointHit}    & Debug except. / Page fault except. & \\
      \texttt{WatchpointHit}    & Page fault except. & \\
      \texttt{FunctionEntry}    & Breakpoint on function entry point & \\
      \texttt{FunctionExit}      & Breakpoint on return address & \\
      \texttt{SyscallEntry}  & Breakpoint on syscall entry point  & \\
      \texttt{SyscallExit}   & Breakpoint on return address & \\
      \texttt{IOOperationPort}  & Port read/write & $\surd$ \\
      \texttt{IOOperationMmap}  & Watchpoint on device memory & \\
      \bottomrule
    \end{tabular}
  }
  \caption{Techniques for tracing events}
  \label{tab:exits}
\end{table}

\T~\ref{tab:exits} describes the technique used to trace all the events
currently supported by the framework. Low-level events (those with a mark in
the last column) correspond directly to exits to root mode (\textit{e.g.,}
\texttt{Exception}). Other events are traced through the aforementioned ones
(\textit{e.g.,} \texttt{Breakpoint\-Hit}), and others again are traced through the latter
(\textit{e.g.,} \texttt{FunctionEntry}).

Events that can be traced directly through the hardware are process switches,
exceptions, interrupts, and port-based I/O operations. All these events exit
conditionally: they exit to root mode only when requested and can have optional
exit conditions to limit exits to particular situations. The remaining of this
section presents how we developed the primitives for tracing higher-level events
starting from the aforementioned low-level ones.

Breakpoints and watchpoints are two of the most complicated events to
implement. Modern CPUs provide hardware facilities to realize efficient and
transparent breakpoints and watchpoints. Unfortunately, hardware-assisted
breakpoints and watchpoints are limited in number (only 4) and shared between
non-root and root mode. Therefore, they cannot be used simultaneously by
the analyzed system and by the framework.
The solution we adopt to allow an arbitrary number
of breakpoints is to use \emph{software breakpoints}. A software breakpoint is
a one-byte instruction that triggers a breakpoint exception when
executed. Software breakpoints are enabled by replacing the byte at the address
on which we want the breakpoint with the aforementioned instruction. When the
breakpoint is hit, the original byte is restored and the event is notified to
the tool. If the breakpoint is not persistent the execution of the system is
resumed. Otherwise the instruction is emulated and then the breakpoint is set
again. Clearly, this approach to breakpoints is not transparent for the
analyzed system. However, it is very efficient. An alternative and transparent
approach is to use the same technique we use for watchpoints, as described in
the next paragraph. Our framework supports both approaches.

The approach used in our framework to implement software watchpoints is based
on protecting the memory locations from any access via hardware (or just from
write accesses, depending on the type of watchpoint), such that any access
results in an exception~\cite{vasudevan05:vampire}. More precisely, since the
finest level of protection offered by the hardware is at the page level, we
mark the page containing the address on which we want to set the watchpoint as
``non-present''. Any future access to this page will result in a page fault
exception that will be intercepted by our framework. The framework analyzes the
exception and checks whether the accessed address corresponds to the address
with the watchpoint. If the watchpoint is hit, the framework delivers the event
to the analysis tool, otherwise it emulates the instruction, and then resumes
the normal execution of the guest. Emulation is necessary to execute the faulty
instruction manually. Indeed, to prevent a second fault, the original
permission of the memory page accessed by the instruction must be restored before
executing the faulty instruction. After the execution of the instruction, the
page must be marked again as ``non-present'' to catch future accesses.

Other higher-level events, such as function and system call entries and exits,
are traced through breakpoints. When the analysis tool requests the framework
to monitor a certain function, the framework sets a breakpoint on the address
of the entry point of the function. Later, when a breakpoint is hit, the
framework checks whether the hit breakpoint corresponds to a function entry
point and, if so, it delivers the appropriate event (\textit{i.e.,}
\texttt{FunctionEntry}) to the analysis tool. Function exits, instead, are
traced by setting a breakpoint on the return address. The framework discovers
the return address by setting a breakpoint on the function entry and by
inspecting the stack frame of the function when the breakpoint on the entry
point is hit. A similar approach is used for tracing system calls entries and
exits.

The approach for tracing function calls and returns just described allows to
trace specific functions, whose names or addresses are supplied by the
tool. The tracing of all function calls and returns is instead more complicated
because it is not possible to know \emph{a priori} the addresses of all
functions' entry points. The solution in this case is to perform a static
analysis to identify the addresses of all functions' entry points (\textit{e.g.,} by
recognizing function prologues). This feature is still not available in our
current implementation of the framework. Nevertheless, if needed, the static
analysis could be performed directly in the tool. The tracing of all system
calls is instead much easier, since they are all invoked through a common
gate. The solution we adopt is to put a breakpoint on the entry point of the
system call gate~\cite{dinaburg:08:ether}.

Beside execution tracing facilities, the framework also exposes to analysis
tools the possibility of intercepting I/O operations with hardware peripherals.
Software can interact with hardware devices through hardware I/O ports, or it
can leverage memory-mapped I/O. In the first case, VMX allows to intercept the
operation without any effort: the framework simply configures the execution
control fields such that all the interactions with the specific hardware ports
trigger an exit to root mode; when such an exit occurs, the framework notifies
the tool by means of a \texttt{IOOperationPort} event. However, for performance
reasons, modern peripherals typically resort to memory-mapped I/O. In this
case, read and write operations do not involve any hardware port, as they are
performed directly on memory. To intercept such operations we set a watchpoint
on the appropriate memory region. Thus, when an access to it is
detected, the framework delivers a \texttt{IOOperationMmap} event to the tool.

\subsection{State Inspection and Manipulation}
\label{sec:mmu}

Several situations require to access the state of the guest system in order to
inspect, and optionally manipulate, both the registers of the CPU and the
memory. As an example, the framework could need to read the return address of a
function from the stack, to access the parameters of a system call from the
processor registers, or to insert a breakpoint into the address space of a
particular process. Similarly, the analysis tool might need to extract data
from the memory of the guest. 

The inspection and manipulation of CPU registers is a straightforward
activity. These information are saved during an exit and restored before an
entry. Thus, the inspection and manipulation of registers merely consists of
reading or writing the VMX guest state (or the memory of the framework,
depending on the type of register).

Inspection and manipulation of memory locations is much more complex. When
paging is enabled, virtual addresses are translated by the hardware into
physical addresses according to the content of the page table and direct
physical addressing is not possible. Each process has its own page table;
therefore, different processes have different virtual-to-physical mappings and
a process cannot access the memory of the others. The framework is isolated
from the guest using the same approach and thus it has its own page table and
its own mapping. Consequently, the framework cannot directly access memory
locations of guest processes. Moreover, inspection is complicated by the fact
that page tables cannot be traversed via software (but only via hardware): the
page table is a multilevel table and pointers to lower levels are physical. To
overcome this problem we have developed a specific, OS-independent, algorithm
that allows to access an arbitrary virtual memory location of an arbitrary
process. The core of the algorithm is a primitive that allows to access
arbitrary physical memory locations. This is accomplished by mapping a given
physical address $p$ to an unused virtual address $v$ in the page table of the
framework, and subsequently by accessing $v$. Then, using this primitive, the
algorithm can traverse the page table of a process of the guest via software by
iteratively mapping the physical addresses stored in the table.

%
The framework exposes memory inspection and manipulation facilities, based on
the aforementioned algorithm, to the analysis tools through two API functions:
\texttt{Guest\-Read(\textit{p},\-\textit{a},\-\textit{n})} and
\texttt{Guest\-Write(\textit{p},\textit{a},\textit{data})}. The former reads
\texttt{\textit{n}} bytes starting from virtual address \texttt{\textit{a}} of
process \texttt{\textit{p}}; the latter writes the content of buffer
\texttt{\textit{data}} into the address space of process \texttt{\textit{p}},
starting from virtual address \texttt{\textit{a}}. By default, to preserve the
integrity of the guest, all \texttt{GuestWrite} operations are forbidden. On
top of this functions we have built higher-level ones that facilitates the
extraction of functions' arguments, null terminated strings, and to disassemble
code. 


\subsection{Tool Isolation}
\label{sec:isolation}

To be able to use our infrastructure on a production system, it is essential to
guarantee that any defect in the analysis tool will not affect the stability of
the analyzed system and of the framework. At this aim, the framework controls
the execution of the analysis tool and, if any anomalous behavior is observed,
the whole infrastructure is automatically unloaded.

As we outlined at the beginning of this section, even if the analysis tool is
executed in VMX root mode, it is still constrained into a less privileged
execution mode than the framework. Thus, any operation the tool performs on the
guest must be mediated by the framework. This is exactly what happens in
traditional operating systems: a user-mode process cannot access directly the
resources of the operating system, nor those of other user-mode processes, and
any action it performs outside its address space must be mediated by the
kernel. Similarly in our context, to perform an operation on the guest system,
the tool must use the programming interface offered by the
framework. 

In the default configuration, the framework does not allow a tool to access in
write-mode to the state of the guest.  However, there is still the possibility
that the execution of an instruction of the tool raises an unexpected exception
(\textit{e.g.,} a page fault on memory access, or a general protection fault). When such
an event occurs, the framework has no way to handle the anomalous situation and to
allow the tool to continue its execution. The only viable approach that also
preserves the integrity of the guest system is to terminate the analysis tool
and to remove the framework.
At this aim, the solution we adopt is to intercept unexpected exceptions
through the custom interrupt descriptor table (IDT) installed when launching
VMX modes. The IDT receives the trap, and delivers it to the trap gate that
eventually unloads the framework. Another problem that might arise with a buggy
analysis tool is non-termination: if the analysis tool entered an infinite
loop, the guest system would never be resumed. To prevent this problem we added
to the framework a minimalistic watchdog and set a time limit on the execution
of the tool. The limit is not on the whole execution time of the tool, but
rather on the execution time to handle an event. Thus, the analysis tool could
potentially be run forever, but with the guarantee that the execution of the
analyzed system will be resumed within the specified time limit. At this aim,
before delivering an event to the analysis tool, the framework resets a
timer. Then, while the tool handles the event, the framework periodically
regains the control of the execution and checks whether the time limit has been
exceeded. To do that the framework registers, in the IDT, a custom interrupt
handler to handle timer interrupts and programs the interrupt controller to
deliver only timer interrupts (that is necessary to prevent the framework to
consume interrupts for all the other devices). Before returning to non-root
mode, the framework reprograms the interrupt controller to deliver all the
interrupts to the analyzed system.

\subsection{OS-dependent Interface}
\label{sec:osdependent}

Our framework provides a general programming interface completely
independent from the operating system running inside the
guest. However, in many cases some OS-specific facilities can ease the
analysis of the guest. As an example, the only OS-independent manner
to identify a process is by means of the base address of its page
table (typically stored inside the \texttt{cr3} CPU
register). However, it is quite awkward to refer to processes using
page table base addresses, and it is more natural to identify a
process through its process id (PID) or through the name of the
application it executes.

\begin{table}[tbp]
  \centering\scriptsize
  \begin{tabular}{lp{5.5cm}}
    \toprule
    \textbf{Name} & \textbf{Description} \\
    \midrule
    \texttt{GetFuncAddr(\emph{n})}  & Return the address of the function \emph{n} \\
    \texttt{GetFuncName(\emph{a})}  & Return the name of the function at address \emph{a} \\

    \texttt{GetProcName(\emph{p})}  & Get the name of process with page directory base address \emph{p} \\
    \texttt{GetProcPID(\emph{p})}   & Get the PID of process with page directory base address \emph{p} \\
    \texttt{GetProcLibs(\emph{p})}  & Enumerate the dynamically linked libraries loaded into process \emph{p} \\
    \texttt{GetProcStack(\emph{p})} & Get the stack base for process \emph{p} \\
    \texttt{GetProcHeap(\emph{p})}  & Get the heap base for process \emph{p} \\
    \texttt{GetProcList()}          & Enumerate processes \\
    \texttt{GetDriverList()}        & Enumerate device drivers \\
    \bottomrule
  \end{tabular}
  \caption{OS-dependent API}
  \label{tab:osdep}
\end{table}

The OS-dependent interface we provide leverages virtual machine introspection
techniques~\cite{garfinkel03:vmi} to analyze the internal structures of the
guest operating system to translate OS-independent information (\textit{e.g.,} process
with page table base address \texttt{0x13cdc000}) into something more
user-friendly (\textit{e.g.,} process \texttt{notepad.exe}). Moreover, using debugging
symbols, the framework allows to resolve symbols' names and addresses (\textit{e.g.,}
functions and global variables). In this way, a tool can ask to interrupt the
execution of the guest when function \texttt{NtCreateFile} is invoked, instead
of referencing the function through its address. Similarly, when a function is
invoked, it is possible to inspect its call-stack and to resolve the name of the
caller functions and even recover the libraries to which the various functions
belong to. Some of the OS-dependent functionalities provided are summarized in
\T~\ref{tab:osdep}.

In case the guest operating system is not supported, the OS-dependent module is
disabled, and only OS-inde\-pen\-dent functionalities are available. Our
current implementation offers an OS-de\-pen\-dent interface only for the Windows XP
operating system.
 

\section{Applications}

In this section we present \hdbg, an interactive kernel debugger for Microsoft
Windows XP we built on top of our framework. In our strive to contribute to the
open source community, we released the code of \hdbg, along with the code of
the framework, under the GPL (v3.0) license. The code is available at the
following address:
\begin{center}
  \url{http://security.dico.unimi.it/hyperdbg/}
\end{center}
The section also discusses other possible applications that could be
constructed using our framework.

\subsection{HyperDbg}
\label{sec:hyperdbg}

\hdbg{} is an interactive kernel debugger we developed on top of our analysis
framework. It offers all the features commonly found in kernel-level debuggers
but, being completely run in VMX root mode, it is OS-independent and grants
complete transparency to the guest operating system and its applications. The
debugger provides a simple graphical user interface to ease the interaction
with the user. This interface is activated in two circumstances: (\textsc{i})
when the user presses a special hot-key or (\textsc{ii}) when the debugger
receives the notification for an event that requires the attention of the user
(\textit{e.g.,} when a breakpoint is hit). From this interface the user interacts with
the debugger and can perform several operations, including setting breakpoints
and watchpoints, tracing functions and system calls, and inspecting and
manipulating the state of the guest (since all interactive debuggers allow to
modify the state of the debuggee, we decided to enable write access to the guest
as well).

\begin{figure}[tbp]
  \centering
  \includegraphics[width=8.4cm]{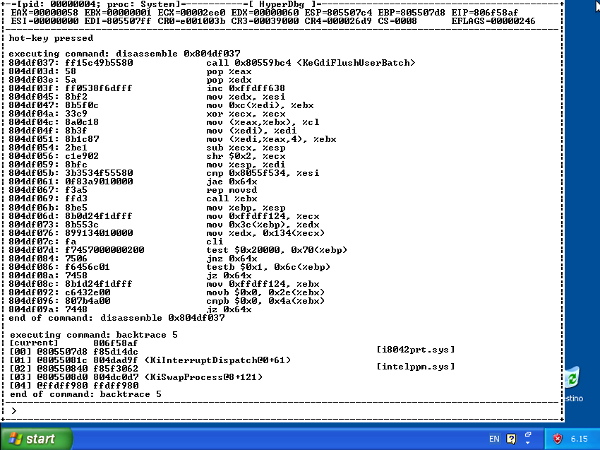}
  \caption{\hdbg{} in action}
  \label{fig:screenshot}
\end{figure}

\F~\ref{fig:screenshot} shows \hdbg{} in action\footnote{The screenshot was
  taken using our development environment based on an Intel x86 emulator
  supporting extensions for virtualization (\textit{i.e.,} BOCHS).}. In particular, the
figure shows the debugger notifying the event that interrupted the execution of
the analyzed system, displaying a fragment of the code of the process currently
running in the analyzed system and displaying a ``backtrace'' of the function
calls that are currently active. Additionally, the debugger displays
information about the status of the registers at the time the event occurred
(in the case of the figure the event is the pressure of the hot-key). To
facilitate the analysis, the debugger leverages OS-dependent
information. For example, the screenshot in
Figure~\ref{fig:screenshot} shows that the debugger resolved the ID and
the name of the process in a MS Windows XP guest, by knowing how the
process table is managed by the operating system.

It is worth pointing out that \hdbg{} can be used to debug \emph{any} piece of
code of the guest system, including critical components such as the process
scheduler, or interrupt and exception handlers. Indeed, \F~\ref{fig:screenshot}
shows that the guest operating system has been stopped while executing the PS/2
keyboard/mouse driver (\texttt{i8042prt.sys}). Thanks to the fact that the
framework on which the debugger is built on is completely transparent to the
analyzed system, the user can use the keyboard to interact with the debugger
even though the keyboard driver of the guest is being debugged.

\hdbg{} consists of less than 1600 lines of code: $\sim$25\% of the code
implements the graphical interface, $\sim$23\% of the code provides the
facilities required for keyboard-based user interaction, and the remaining
$\sim$52\% is responsible for handling events and for all the other
interactions with the framework. Note that certain functionalities (\textit{e.g.,}
disassembling a code region) are implemented directly in the framework since,
most likely, they will be used for other types of analysis as well. The
framework is about four times bigger than the debugger (without considering the
disassembly module embedded in the framework, as it is based on an off-the-shelf
disassembler). We believe these numbers are very significant. The number of
lines of code we had to write to implement \hdbg{} clearly witnesses that
complex analysis tools like an interactive kernel debugger are straightforward
to implement using our framework.

The remaining of this section describes how we used the facilities of the
framework to implement the user interface and the component to receive commands
from the user. 

\paragraph{User Interface}

Although the graphical user interface of the debugger is rough, its
implementation is very challenging. The reason of the complexity is the fact
that we cannot rely on any high-level graphical facility available in the
analyzed system to render the interface. Such approach would be too OS-depended
and not transparent at all. The lack of graphical primitives obliged us to
interact directly with the video card. The video memory is mapped at a fixed
address in the guest and thus unmodified inspection and manipulation API (i.e.,
\texttt{GuestRead} and \texttt{GuestWrite}) can be used by the debugger to
render the interface. Note that this approach is not dependent on the OS nor on
the hardware. We developed a small video library that provides basic graphical
functionalities and translates our requests into data that are written directly
in the memory of the video card. Before rendering the graphical interface to
the screen, the debugger backups the content of the video memory and restores
the content right before resuming the execution of the analyzed
system.

\paragraph{User Interaction}
User interaction is keyboard-based. When in non-root mode, the user can switch
into \hdbg{} by pressing a hot-key. Then, in root mode the user can control the
debugger. For these reasons, \hdbg{} must be able to intercept keystrokes both
in root and non-root mode. 
To intercept keystrokes in non-root mode we monitor all the read operations
from the hardware I/O port devoted to the keyboard. In other words, \hdbg{}
registers to the core for all the \texttt{IOOperationPort} events that satisfy
the event condition \texttt{\emph{port}=KEYBOARD\_PORT \&\&
  \emph{access}=read}. When such operation is detected, \hdbg{} checks whether
the key pressed corresponds to the hot-key that enables the debugger. If the
key pressed matches the hot-key the debugger pops up the graphical interface
and waits for commands. Otherwise, the debugger passes the keystroke to the
analyzed system such that the latter will continue its execution as if the
keystroke were read directly from the keyboard. Keyboard handling in root mode
is done by polling the keyboard hardware I/O port. Since direct access to I/O
ports is not permitted to any analysis tool, the debugger relies on a 
API function exported by the framework which mediates all accesses to I/O ports
and allows (if the permission is granted at compile time) certain analysis
tools to read data from certain I/O ports.

\subsection{Other Possible Uses of the Framework}
\label{sec:others}

\hdbg{} demonstrates that our framework is very versatile and that enables new
opportunities for dynamic analysis and we will explore in our future research. 

An interesting extension of \hdbg{} will be the support for kernel-level
omniscent debugging. Omniscent debugging allows developers to inspect the
status of their programs in past execution instants, in order to detect the
cause of a failure without the need to run the target program multiple
times~\cite{odb}. \hdbg{} could be extended to allow a user to record and
inspect the values a memory location stored during the time, and the exceptions
and interrupts occurred. Such a feature would ease a user to discover when a
memory location of the kernel gets corrupted and which instruction is
responsible for the corruption. Moreover, the ability to log asynchronous
events, such as interrupts, would allow to spot defects connected to
non-deterministic behaviors of the analyzed system. Our framework already
offers all the necessary facilities for this kind of debugging: exception and
interrupts can be traced natively by the framework and memory accesses can be
traced using watchpoints.

Another interesting application of our framework will be dynamic aspect-oriented
programming of operating system kernels. As discussed in \S~\ref{sec:related},
several approaches have been proposed to apply AOP to kernels. The main
advantage offered by our framework over the approaches proposed so far is
that it does not require any modification of the source code of the
kernel, nor any modification of the image in memory of the kernel. Moreover, our
framework protects the running kernel from defects in the woven code. One
approach to facilitate the use of such technology would be to provide
programmers a source-to-source translator, to translate aspect
oriented code written in languages like AspectC~\cite{coady:2001:aspectc} into
C code that uses the API offered by our framework. In particular, the
translator would be responsible for translating pointcuts into API calls to
trace the corresponding events, using advices as events handlers, and for
translating all pointer dereferences into calls to inspection API to read the
memory of the guest.

\section{Conclusions}
\label{sec:conclusions}

We proposed a framework to perform complex run-time analyses of both
system- and user-level code on commodity production systems. The
framework exposes an API that eases the development of analysis tools
on its top. The approach we described leverages hardware extensions
for virtualization available on modern processors to overcome the
limitations that affect existing approaches for the analysis of
system-level code. In particular, the solution we proposed does not
require to recompile or reboot the target system, it is not invasive,
it is almost completely OS-independent, and it guarantees that a
defect in an analysis tool cannot damage the framework itself nor the
analyzed system. To demonstrate its potentials, we developed \hdbg, an
interactive kernel-level debugger for Microsoft Windows XP. \hdbg{}
and the framework have been released as an open source package.

\section*{Acknowledgments}
\label{sec:ack}
This research has been partially funded by the European Commission,
Program IDEAS--ERC, Pro\-ject 227977 SMSCom and by the Italian Ministry
of Education, Universities and Research, Program PRIN--2008.



\bibliographystyle{abbrv}
\bibliography{biblio}

\end{document}